\documentclass[twocolumn,prl,secnumarabic,amssymb,amsmath,nobibnotes,aps,showpacs]{revtex4}
\usepackage{graphics}
\usepackage{graphicx}

\usepackage[usenames,dvipsnames]{color}

\begin{document}

\title{A Dual-Species Bose-Einstein Condensate of $^{87}$Rb and $^{133}$Cs}

\author{D. J. McCarron, H. W. Cho, D. L. Jenkin, M. P. K\"oppinger, and S. L. Cornish}
\affiliation{Department of Physics, Durham University, Rochester Building, Science Laboratories, South Road, Durham DH1 3LE, United Kingdom}

\begin{abstract}
We report the formation of a dual-species Bose-Einstein condensate of $^{87}$Rb and $^{133}$Cs in the same trapping potential. Our method exploits the efficient sympathetic cooling of $^{133}$Cs via elastic collisions with $^{87}$Rb, initially in a magnetic quadrupole trap and subsequently in a levitated optical trap. The two condensates each contain up to $2\times10^{4}$ atoms and exhibit a striking phase separation, revealing the mixture to be immiscible due to strong repulsive interspecies interactions. Sacrificing all the $^{87}$Rb during the cooling, we create single species $^{133}$Cs condensates of up to $6\times10^{4}$ atoms.
%
\end{abstract}


\maketitle

Quantum degenerate mixtures \cite{Myatt1997,Stenger1998,Truscott2001,Hadzibabic2002,Roati2002,Modugno2002,Silber2005,Zaccanti2006,Ospelkaus2006,Papp2008,Taglieber2008,Tey2010} of atomic gases are currently the subject of intensive research, exhibiting rich physics inaccessible in single species experiments. Phase separation of the constituent gases is a particularly dramatic example \cite{Stenger1998,Zaccanti2006,Ospelkaus2006,Papp2008}. In an optical lattice, the myriad of quantum phases for two different atomic species \cite{Altman2003,Kuklov2004} goes far beyond the seminal observations of the superfluid to Mott-insulator transition \cite{Greiner2002} and offers many intriguing applications. Amongst these applications, the creation of degenerate samples of heteronuclear ground state molecules is of paramount interest \cite{Carr2009}. Possessing long-range, anisotropic dipole-dipole interactions, such molecules offer a wealth of new research avenues \cite{Lahaye2009}, including quantum simulation and computation \cite{Micheli2006,Demille2002}. At present the most promising route to these riches exploits a mixed species quantum gas in which the constituent atoms are associated into ground state molecules \cite{Damski2003}. Remarkably, using a combination of magneto-association on a Feshbach resonance \cite{Kohler2006} followed by optical transfer to the rovibrational ground state, this association can be achieved with near unit efficiency, as demonstrated using KRb \cite{Ni2008} and Cs$_2$ \cite{Danzl2008}.


A quantum degenerate mixture of $^{87}$Rb and $^{133}$Cs would provide an ideal gateway into the realm of dipolar molecular quantum gases. Both species have been separately condensed and the intraspecies two-body interactions are now well understood \cite{Marte2002,Chin2004}. Numerous interspecies Feshbach resonances have been observed \cite{Pilch2009}, permitting, in principle, the control of the interspecies interactions and the creation of RbCs Feshbach molecules. Crucially, the RbCs molecule in the rovibrational ground state is stable against atom exchange reactions of the form RbCs + RbCs $\rightarrow$ Rb$_2$ + Cs$_2$, unlike the KRb molecule \cite{Zuchowski2010}. However, associated with the rich Feshbach spectrum in $^{133}$Cs are high two- and three-body inelastic collision rates which complicate the evaporative cooling. Indeed, to date Bose-Einstein condensation of $^{133}$Cs has required sophisticated laser cooling  techniques, and has been achieved by only two groups \cite{Weber2003,Hung2008}.

In this Letter we demonstrate the production of a dual-species Bose-Einstein condensate of $^{87}$Rb and $^{133}$Cs, thereby realizing the perfect starting conditions for the creation of a quantum gas of dipolar RbCs molecules. Our method uses simple established techniques and exploits the efficient sympathetic cooling of $^{133}$Cs via elastic collisions with $^{87}$Rb, initially in a magnetic quadrupole trap and subsequently in a levitated optical trap. The method can easily be adapted to produce large single species $^{133}$Cs condensates, offering an alternative route to quantum degeneracy for this atom.
 %

\begin{figure}[t]
\centering
\includegraphics[width=\columnwidth]{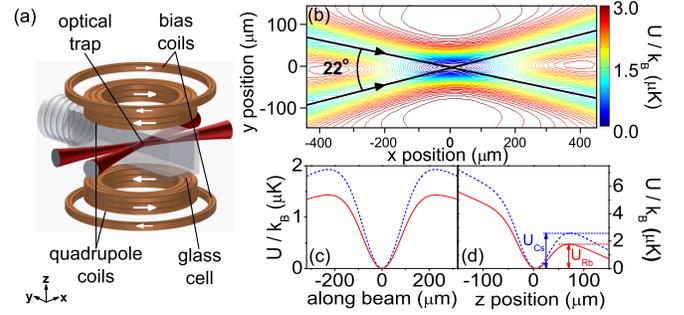}
\caption{(Color online) (a) Geometry of the levitated crossed dipole trap used in the experiment. (b-d) Typical trapping potential employed towards the end of the evaporation sequence (the power of both beams is 120~mW, the magnetic field gradient is 35~G/cm and the bias field is 22.4~G). The contour plot in (b) shows the potential for $^{87}$Rb. The cut along $z$ in (d) highlights the tilting of the trap.} \label{fig:Potential}
\end{figure}


At the heart of our experiment is a levitated crossed dipole trap \cite{Jenkin2011} located at the centre of an ultrahigh vacuum (UHV) glass cell (Fig.\,\ref{fig:Potential}). Two 1550~nm beams derived from a 30~W fibre laser (IPG ELR-30-LP-SF) are focussed to waists of $\sim60~\mu$m and intersect $\sim80$~$\mu$m below the unbiased field zero of a magnetic quadrupole potential. The power in each beam is independently controlled using acousto-optic modulators. The magnetic potential generated by the quadrupole and bias coils allows atoms in the trap to be levitated against gravity. Both species can be levitated simultaneously due to the fortuitous coincidence of the magnetic moment to mass ratios for the $^{87}$Rb $|F=1, m_F=+1 \rangle$ and $^{133}$Cs $|3, +3 \rangle$ states. When loading the trap the power in each beam is set to 6~W. This creates a trap $90~\mu$K ($125~\mu$K) deep for $^{87}$Rb ($^{133}$Cs). The difference in the trap depths arises from the different polarizabilities at 1550~nm and makes the trap potential suitable for sympathetic cooling of $^{133}$Cs. For this loading potential, the axial and radial trap frequencies for $^{87}$Rb ($^{133}$Cs) are $\nu_{z} = 135~$Hz (127~Hz) and $\nu_{r} = 680~$Hz (640~Hz) respectively.


To load the dipole trap, we first collect ultracold mixtures of $^{87}$Rb and $^{133}$Cs in a two-species magneto-optical trap (MOT) in the UHV cell \cite{Harris2008}. Up to $3\times10^{8}$ $^{87}$Rb atoms are collected. The number of $^{133}$Cs atoms in the MOT is stabilized between $5\times10^{5}$ and $3\times10^{8}$ by monitoring the MOT fluorescence and actively controlling the repump light level. This allows precise control of the species ratio in the mixture. Following a compressed MOT and molasses phase the atoms are optically pumped into the $^{87}$Rb $|1,-1 \rangle$ and $^{133}$Cs $|3,-3 \rangle$ states and the mixture is loaded into a magnetic quadrupole trap. The trap gradient is adiabatically increased from 40 to 187~G/cm, prior to using forced RF evaporation to cool the $^{87}$Rb gas. To probe the mixture, the trap is switched off and resonant absorption images are captured for both species in each iteration of the experiment using a frame transfer CCD camera \cite{Harris2008}. The trap depth set by the RF frequency is three times greater for $^{133}$Cs than for $^{87}$Rb and therefore allows the selective removal of hot $^{87}$Rb atoms. As the $^{87}$Rb gas is cooled we observe efficient sympathetic cooling of $^{133}$Cs via interspecies elastic collisions \cite{Cho2011}. The efficiency, $\gamma$, ranges from $5.2(4)$ to $11.3(4)$ depending on the initial number of $^{133}$Cs atoms (the efficiency is defined as $\gamma = -\frac{log(D_{0}/D)}{log(N_{0}/N)}$, where $D_{0}$ ($N_{0}$) and $D$ ($N$) are the initial and final phase-space densities (numbers) respectively). The high efficiency of the sympathetic cooling implies a large interspecies elastic collision rate. This is consistent with the latest theoretical refinement of the interspecies collision potential \cite{HutsonPC} which shows excellent agreement with the published Feshbach spectrum \cite{Pilch2009,Lercher2011} and indicates a background interspecies scattering length of $\approx650~a_{0}$ ($a_{0}=0.0529~$nm) for the states used in this work. We continue this cooling until Majorana losses begin to limit the lifetime of the sample. At this stage the crossed dipole trap is loaded via elastic collisions by simply reducing the magnetic field gradient to 29~G/cm, just below the 30.5~G/cm (31.1~G/cm) required to levitate $^{87}$Rb ($^{133}$Cs) \cite{Porto2009}. At the same time the final RF frequency is adjusted to control the composition of the mixture loaded into the dipole trap.

\begin{figure}[t]
\centering
\includegraphics[width=\columnwidth]{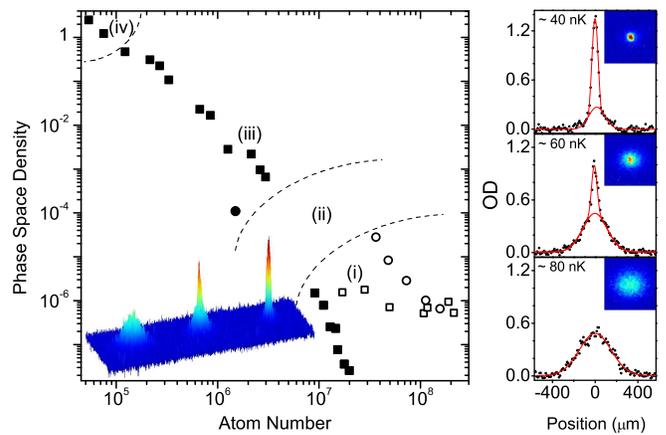}
\caption{(Color online) Trajectory to $^{133}$Cs BEC via sympathetic cooling with $^{87}$Rb (squares). Open (closed) symbols show $^{87}$Rb ($^{133}$Cs) data. Circles show single species data for comparison. The trajectory gradient indicates the cooling efficiency. The different regions show: (i) RF evaporation in the magnetic quadrupole trap. (ii) Dipole trap loading and internal state transfer. (iii) Evaporation by reducing the dipole trap beam powers. (iv) Evaporation by trap tilting. Inset: 3D rendering of the density distribution as the cloud is condensed. Right: absorption images ($350\times350~\mu$m) and cross sections of the cloud near the BEC phase transition.} \label{fig:Trajectory}
\end{figure}

To create $^{133}$Cs condensates a mixture containing $2.1(1)\times10^{8}$ $^{87}$Rb and $2.0(2)\times10^{7}$ $^{133}$Cs atoms is loaded into the quadrupole trap. The evaporation efficiency for $^{87}$Rb alone in the magnetic trap is 2.6(2) (\Large{$\circ$} \normalsize in Fig. \ref{fig:Trajectory} (i)). For the two species case the efficiencies are 0.4(1) and 5.2(4) for $^{87}$Rb and $^{133}$Cs respectively, as we sacrifice $^{87}$Rb to sympathetically cool $^{133}$Cs (\small{$\square$} \normalsize and \small{$\blacksquare$} \normalsize in Fig. \ref{fig:Trajectory} (i)). As the dipole trap is loaded the RF frequency is reduced to optimise the transfer of $^{133}$Cs whilst at the same time removing all of the $^{87}$Rb. The resulting loading is highly efficient with $\sim50~\%$ of the $^{133}$Cs atoms transferred into the dipole trap at a temperature of 10(1)~$\mu$K. The phase-space density is $6.6(2)\times10^{-4}$, representing more than a factor of 400 increase compared to the final value in the quadrupole trap. This increase results from an effect analogous to the use of a dimple potential \cite{Weber2003}. Adiabatic expansion of the trap volume cools the cloud in the quadrupole potential and elastic collisions transfer some atoms into the tighter harmonic optical potential \cite{Porto2009}. The remaining atoms are lost from the quadrupole trap reservoir, removing energy in the process. We believe the enhanced loading of $^{133}$Cs in the presence of $^{87}$Rb compared to the $^{133}$Cs alone case (\Large{$\bullet$} \normalsize in Fig. \ref{fig:Trajectory}) is again consistent with the existence of a large interspecies scattering length \cite{HutsonPC,Lercher2011}.

To proceed we transfer the $^{133}$Cs atoms to the $|3,+3\rangle$ state with $\sim95$~\% efficiency using RF adiabatic rapid passage as a bias field is turned on to $\sim22$~G in 18~ms \cite{Jenkin2011}. This transfer eliminates inelastic two-body losses which have plagued previous attempts to cool $^{133}$Cs to degeneracy in magnetic traps \cite{GueryOdelin1998}. In addition, at this bias field the ratio of elastic to inelastic three-body collisions is favorable for evaporative cooling \cite{Kraemer2004}. Operating in the vicinity of the broad zero-crossing in the scattering length near 17~G gives precise control of the elastic collision rate. We observe efficient evaporation for fields between 21 and 24~G, and typically operate at 22.4~G where $a_{\rm{Cs}} = 280~a_{0}$. The majority of the evaporation is performed by reducing the beam powers to 48~mW over 9.5~s. The final evaporation to degeneracy is implemented by increasing the magnetic field gradient to 35.4(1)~G/cm over 2.0~s. This tilts the trapping potential (Fig.\,\ref{fig:Potential} (d)), thereby lowering the trap depth whilst leaving the trap frequencies largely unchanged \cite{Hung2008}. With this approach we achieve an evaporation efficiency of 2.1(1) for $^{133}$Cs alone in the dipole trap and produce pure single species  condensates of up to $6.2(1)\times10^{4}$ atoms. Fig. \ref{fig:Trajectory} summarizes the trajectory to $^{133}$Cs BEC. We are not able to produce condensates by loading only $^{133}$Cs into the magnetic trap due to the poor transfer into the dipole trap (\Large{$\bullet$} \normalsize in Fig. \ref{fig:Trajectory}). In contrast, loading only $^{87}$Rb initially we are able to form condensates of up to $1\times10^{6}$ atoms \cite{Jenkin2011}.

\begin{figure}[t]
\centering
\includegraphics[width=\columnwidth]{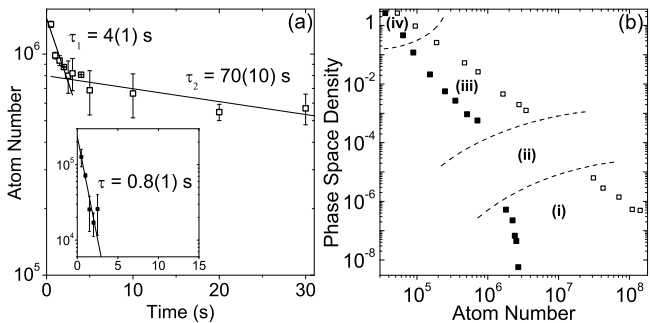}
\caption{(a) Interspecies inelastic losses in the dipole trap. To determine trap lifetimes the majority species ($^{87}$Rb) data are fit with a double exponential function while the minority species ($^{133}$Cs) data shown in the inset are fit with a single exponential function. (b) Trajectory to a dual-species BEC of $^{87}$Rb and $^{133}$Cs. The trajectory is divided into four sections: (i) RF evaporation in the magnetic quadrupole trap. (ii) Dipole trap loading and internal state transfer. (iii) Evaporation by reducing the dipole trap beam powers. (iv) Evaporation by trap tilting. Open (closed) symbols show data for $^{87}$Rb ($^{133}$Cs).}  \label{fig:DualspecBEC}
\end{figure}
To create a degenerate mixture of $^{87}$Rb and $^{133}$Cs we reduce the number of $^{133}$Cs atoms loaded into the magnetic trap to $2.7(1)\times10^{6}$. This reduces the heat load on the $^{87}$Rb during the sympathetic cooling and improves the cooling efficiencies in the magnetic trap to 1.7(1) for $^{87}$Rb and 11.3(4) for $^{133}$Cs. Loading the dipole trap results in a mixture of $2.8(2)\times10^{6}$ $^{87}$Rb and $5.1(3)\times10^{5}$  $^{133}$Cs atoms at a temperature of 9.6(1)~$\mu$K following the spin flip to the $^{87}$Rb $|1,+1 \rangle$ and $^{133}$Cs $|3,+3 \rangle$ states . The initial peak densities for both species are $>10^{13}$~cm$^{-3}$ and we observe very strong interspecies inelastic losses, see Fig. \ref{fig:DualspecBEC}(a). The fast loss of $^{133}$Cs atoms from the trap shown in the inset of Fig. \ref{fig:DualspecBEC}(a) is purely due to inelastic three-body collisions. The initial decrease of the number of $^{87}$Rb atoms includes a significant contribution from plain evaporation from the trap. Three body collisions are most prevalent at the trap centre and cause the coldest atoms to be lost predominantly. This `anti-evaporation' leads to both atom loss and heating and is a big obstacle when trying to produce a dual-species BEC. The heating leads to further loss from the trap through evaporation and complicates the determination of the three-body loss rate. A full analysis of the evolving densities using two coupled differential rate equations \cite{Barontini2009} is therefore not performed. Instead we estimate an upper limit for the interspecies loss rate coefficient of $\sim10^{-25}-10^{-26}$~cm$^{6}$s$^{-1}$ based simply upon the observed lifetime of the minority species and the average initial densities. The three-body loss rate coefficient scales as $L_{3} = 3C\frac{\hbar}{m}a^{4}$, where $C$ is a dimensionless factor between 0 and $\sim$70, $m$ is the reduced mass and $a$ is the scattering length \cite{Bedaque2000}. The current best estimate \cite{HutsonPC,Lercher2011} of the interspecies scattering length, $a_{\rm{RbCs}}\approx650~a_{0}$, would lead to a \emph{maximum} three-body loss rate of $L_{3}\sim3\times10^{-25}~\rm{cm}^{6}/\rm{s}$ in reasonable agreement with our simple estimates.


To combat the strong interspecies inelastic losses the beam powers are reduced to $120$~mW in just 1.0~s, corresponding to a final trap depth of 2~$\mu$K for $^{87}$Rb. Taking into account the adiabatic expansion of the cloud resulting from the commensurate relaxation of the trap frequencies, this corresponds to an effective reduction in the trap depth by a factor of 7. The result is a factor of $\sim3$ reduction in the density and consequently an order of magnitude increase in the lifetime of $^{133}$Cs against interspecies three-body collisions. Overall the conditions for evaporative cooling are improved by this fast reduction in the dipole trap power, despite the associated decrease in the elastic collision rate. Further evaporation is then performed by reducing the beam powers linearly to 48~mW in 2.5~s. Finally by again tilting the trap using the applied magnetic field gradient dual-species condensates are produced in the same trapping potential containing up to $\sim2.0\times10^{4}$ atoms of each species. Similar results have recently been obtained by confining the $^{87}$Rb and $^{133}$Cs atoms in spatially separated optical traps to suppress the interspecies inelastic losses \cite{Lercher2011}. The trajectory to a dual-species BEC is presented in Fig. \ref{fig:DualspecBEC}(b). The efficiencies of the evaporative and sympathetic cooling in the dipole trap are 1.8(2) and 2.8(5) for $^{87}$Rb and $^{133}$Cs respectively. The data show that despite strong interspecies inelastic losses the sympathetic cooling of $^{133}$Cs via $^{87}$Rb in the dipole trap is still more efficient than the direct evaporation of $^{133}$Cs alone (Fig. \ref{fig:Trajectory}).

\begin{figure}[t]
\centering
\includegraphics[width=\columnwidth]{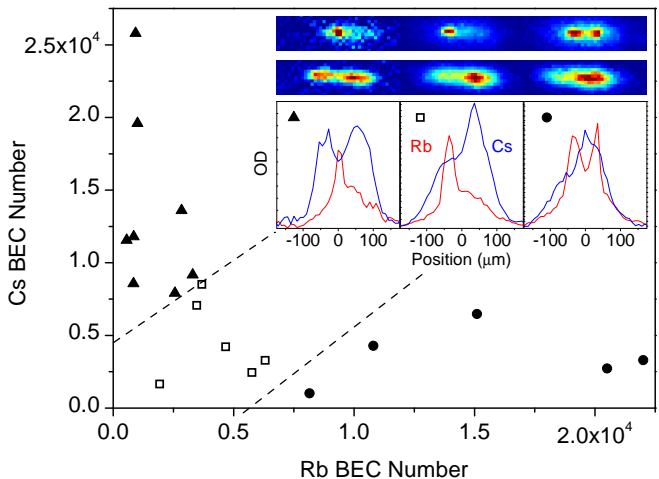}
\caption{(Color online) Immiscibility in a quantum degenerate mixture of $^{87}$Rb and $^{133}$Cs. Three distinct structures are observed ($\blacktriangle$, $\square$ and \normalsize{$\bullet$}\small) correlated with the number of atoms in each condensate.  Absorption images and cross-sections of $^{87}$Rb (red) and $^{133}$Cs (blue) highlight these structures. The optical depth (OD) scales of the cross-sections are 0--1.0 ($\blacktriangle$) and 0--2.6 ($\square$ and \normalsize{$\bullet$}\small). Each image is $350 \times 100~\mu$m.} \label{fig:Immiscible}
\end{figure}

A dramatic spatial separation of the two condensates in the trap is observed, revealing the mixture to be immiscible at 22.4~G (Fig. \ref{fig:Immiscible}). The resulting density profile for each species is a stark contrast to the symmetry of the potential and the density profiles observed for each single species condensate. For an immiscible quantum degenerate mixture, the relative strength of the atomic interactions $\Delta = \frac{g_{\rm{RbCs}}}{\sqrt{g_{\rm{Rb}}g_{\rm{Cs}}}}>1$, where the interaction coupling constant $g_{ij} = 2 \pi \hbar^{2} a_{ij}\Big(\frac{m_{i}+m_{j}}{m_{i} m_{j}}\Big)$ \cite{Riboli2002}. At 22.4~G $a_{\rm{Rb}} = 100~a_{0}$ and $a_{\rm{Cs}} = 280~a_{0}$, so that the observation of immiscibility implies $a_{\rm{RbCs}} > 165~a_{0}$; again consistent with the current best estimate, $a_{\rm{RbCs}}\approx650~a_{0}$ \cite{HutsonPC,Lercher2011}.
The bimodal distributions in Fig. \ref{fig:Immiscible} show that the immiscible behavior is exclusive to the condensed atoms. The immiscible condensates always form one of three structures, either one of two possible symmetric cases (\small{$\blacktriangle$} \normalsize and \Large{$\bullet$} \normalsize in Fig. \ref{fig:Immiscible}) or an asymmetric case (\small{$\square$} \normalsize in Fig. \ref{fig:Immiscible}). Qualitatively, these observations are strikingly similar to the symmetric and asymmetric structures predicted for an immiscible binary condensate based upon numerical simulations of two coupled Gross-Pitaevskii equations \cite{Trippenbach2000}. Experimentally, the structures are correlated with the number of atoms in each condensate. Altering the number of $^{133}$Cs atoms loaded into the trap, allows navigation between the three structures. The strong interspecies losses make the formation of a dual-species condensate with $>2\times10^{4}$ atoms of each species a challenge and lead to the asymptotic nature of the data in Fig. \ref{fig:Immiscible}.

In summary, we have demonstrated the production of a dual-species Bose-Einstein condensate of $^{87}$Rb and $^{133}$Cs in the same trapping potential. Our method uses simple established techniques and exploits the large interspecies scattering length to efficiently sympathetically cool $^{133}$Cs in both the magnetic and optical trapping phases of the experiment. We observe a striking phase separation of the two condensates revealing the mixture to be immiscible due to the strong repulsive interspecies interactions. We believe that the immiscibility aids the formation of the dual-species condensate by suppressing the interspecies inelastic losses as the mixture condenses. In the future, we plan to explore the miscibility by Feshbach tuning both $a_{\rm{Cs}}$ and $a_{\rm{RbCs}}$ in order to test this assertion. Crucially, thermal clouds very close to degeneracy remain miscible, enabling the efficient magneto-association \cite{Kohler2006} of RbCs molecules in future experiments, as the first step towards the creation of a dipolar molecular quantum gas.

DJM and HWC contributed equally to this work. SLC acknowledges the support of the Royal Society. This work was supported by the UK EPSRC and ESF within the EuroQUAM QuDipMol project.


\end{document}